\documentclass[referee]{raa}            % referee version: for submission
\usepackage{bbm}
\usepackage{txfonts}
\usepackage{graphicx,times}             %for PS/EPS graphics inclusion, new
\input{epsf.sty}                        %for PS/EPS graphics inclusion, old
\input{psfig.sty}                       %for PS/EPS graphics inclusion, old

\begin{document}

\title{The first detection of the solar U+III association with an antenna prototype for the future lunar observatory}
%   \subtitle{I. Place Your Subtitle Here}

\volnopage{Vol.0 (200x) No.0, 000--000}      %%preserved for Editor. DOn't remove!
\setcounter{page}{1}          %%starting page, preserved for Editor. DOn't remove!

\author{Lev Stanislavsky\inst{1} \and Igor Bubnov\inst{1,2} \and Alexander Konovalenko\inst{1} \and Peter Tokarsky\inst{1} \and Serge Yerin\inst{1,2}}

\institute{$^1$ Institute of Radio Astronomy, Kharkiv, Ukraine; {\it lev.stanislavskyi@gmail.com}\\
$^2$ V.N. Karazin Kharkiv National University, Kharkiv, Ukraine\\
\vs\no
   {\small Received~~2020 December 11; accepted~~2021 January 29}}

\abstract{We report about observations of the solar U+III bursts on 5 June of 2020 by means of a new active antenna designed to receive radiation in 4-70 MHz. This instrument can serve as a prototype of the ultra-long-wavelength radiotelescope for observations on the farside of the Moon. Our analysis of experimental data is based on simultaneous records obtained with the antenna arrays GURT and NDA in high frequency and time resolution, e-Callisto network as well as by using the space-based observatories STEREO and WIND. The results from this observational study confirm the model of Reid and Kontar (\cite{kontar17}).
}

\authorrunning{Lev Stanislavsky et al.}            %author_head in even pages
\titlerunning{The first detection of the solar U+III association}  % title_head in odd pages

\maketitle
\section{Introduction}
\label{sect:intro}
After a long break, the return of man to the Moon has become extremely important for many reasons. On the one hand, the Moon can serve as an ideal launching pad for building large-scale bases from which future space missions with direct human participation would begin to explore the other inner planets of our solar system, in particular Mars (Witze~\cite{witze19}, Dhingra~\cite{dhingra18}). In the second place, the presence of a unique radio quiet zone on the lunar farside would allow observing very low-frequency emission from many various cosmic objects almost not available for ground-based observations because of ionospheric cutoff as well as intensive artificial and/or terrestrial radio disturbances (Jester and Falcke~\cite{jester09}, Lazio et al.~\cite{lazio11}, Mimoun et al.~\cite{mimoun12}, Zarka et al.~\cite{zarka12} and recent link https://www.space.com/nasa-telescope-far-side-of-moon.html). The making of a large scale radio array is not an easy task even on Earth, and on the other planets and moons this task is a challenge for mankind to build scientific instruments in extremal conditions. The Ukrainian program of lunar explorations by spacecrafts has been suggested by Shkuratov et al.~(\cite{shkuratov2019}), and perspective steps to solar studies with help of radio observations on the lunar far side were considered by Stanislavsky et al.~(\cite{stan2018}). Advancing this program as applied to radio astronomy purposes, we produced an antenna prototype for the future lunar telescope at ultra long wavelengths. It can record radio emission of cosmic objects both independently and as parts of antenna arrays or interferometers. Its important advantage is a feasibility of preliminary approbation on Earth, receiving radio emission in frequencies close to ionospheric cutoff and higher. The present paper is just devoted to real observations as a test of this antenna demonstrating a scope of this instrument for exploring the Universe at ultra long wavelengths in radio emission. Consequently, we observed and study the solar type U radio burst associated with type III bursts that are useful to examine possible models of the event.

\section{Instrument}
\label{sect:facilities}

The great interest in the development of small-sized active dipoles for low-frequency radio astronomy has noticeably intensified theoretical and experimental studies of antenna technology, amplifiers and related components (see Konovalenko et al. ~\cite{konov16} and references therein). Over fifteen years of experience in operating active dipoles in the GURT ground-based radio array showed their reliability and validity of scientific results. These dipoles, being small in size, provide optimal ``radio astronomical sensitivity'', which is determined primarily by the contribution of the amplifier temperature to the noise temperature of the active dipole. In the case of a noise-free amplifier, the noise temperature of the active dipole is equal to the antenna temperature, obtained from observations of the Galactic background radio emission. A good result is considered to achieve the contribution of the amplifier temperature to the noise temperature of the active dipole no more than 10\% - 25\%. In this way, the corresponding studies were carried out to develop a prototype of the ultra-long-wavelength broadband antenna for radio astronomy purposes in which the Moon is an instrument location. 

By numerical simulations of the antenna prototype and performing measurements of its parameters, we have studied the active-dipole antenna of complex geometry, located above partially conductive ground (Tokarsky~\cite{tokar17}). The sketch of this antenna is shown in Fig.~1. Consequently, the computer model of this antenna was developed, which allowed us to obtain its characteristics (impedance, energy parameters, radiation pattern) in the operating frequency domain of 1 to 70 MHz. As a part of these studies, the dipole and the amplifier were designed and manufactured as an active antenna, capable of receiving cosmic radiation in the frequency band 4-70 MHz. The test measurements of the Galactic background radiation were successfully conducted at the Radio Astronomical Observatory of S. Ya. Braude in terrestrial conditions. The antenna design was the following. The dipole arm lengths along the midline (ABC in the frontal projection of Fig.~1) are 2.8 m, the angle $\gamma$ of inclination to the horizontal plane is 45$^\circ$, and the dipole terminals are located at $h$ = 1.7 m above ground. It is made of steel tubes having the diameter of 23 mm. As an antenna amplifier, the low-noise amplifier (LNA) with PHEMT transistors is used (Korolev~\cite{korolev}). The LNA nominal gain is $G_{amp}= 22$ dB, the effective noise temperature is about 50 K, and the 3rd order nonlinear distortion coefficient is more than 30 dB/$\mu$V. The LNA is supplied with the voltage of 5 V at the current consumption of 40 mA.

\begin{figure}
%\centerline{\includegraphics[width=13cm, angle=0]{antenna.eps}}
\centerline{\includegraphics[width=8cm, angle=0]{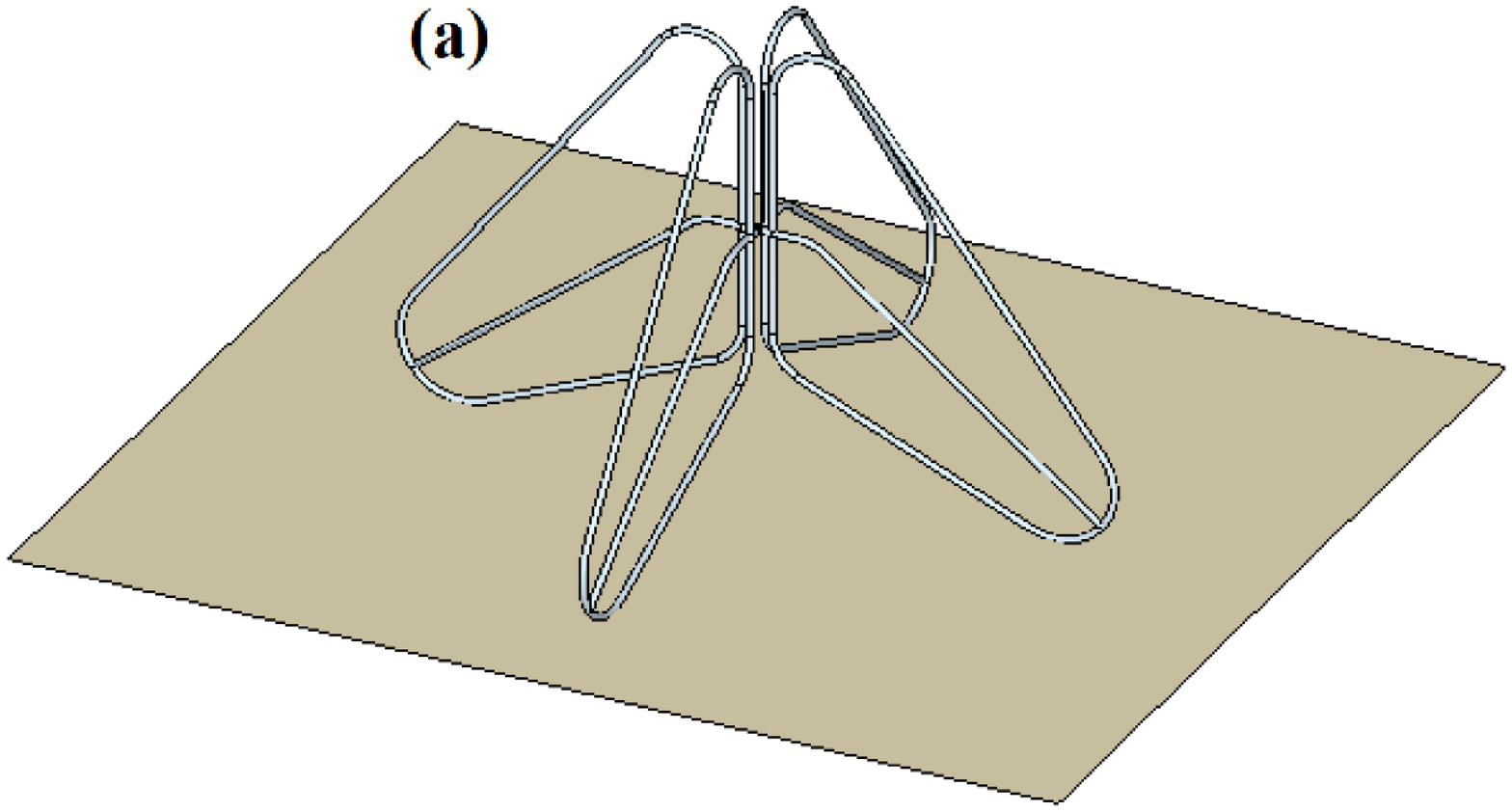}
\includegraphics[width=5cm, angle=0]{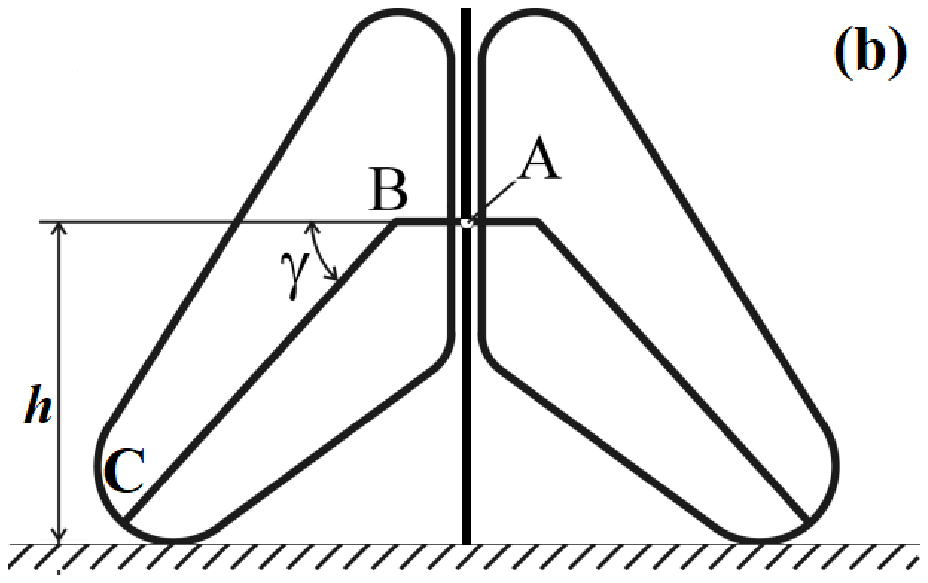}}
\caption{Sketch of the antenna prototype (a) with its frontal projection (b).}
\label{fig0}
\end{figure}

\begin{table}
 \centering
 \caption{Evolution of the active region 12765 (see https://www.spaceweatherlive.com/en/solar-activity/region/12765).}\label{tab1}
 \vspace*{1ex}
 \begin{tabular}{|c|c|c|c|c|}
  \hline
   Data & Number of & Size, & Class & Location, \\
	      & sunspots  & MH    & Magn. & degree    \\
  \hline
  2020/06/03 & 2 & 70 & $\alpha$ & S24E71 \\
  2020/06/04 & 2 & 100 & $\alpha$ & S24E58 \\
	2020/06/05 & 3 & 130 & $\beta$ & S24E44 \\
	2020/06/06 & 5 & 110 & $\beta$ & S22E33 \\
	2020/06/07 & 6 & 100 & $\beta$ & S23E20 \\
	2020/06/08 & 7 & 100 & $\beta$ & S24E07 \\
	2020/06/09 & 4 & 70 & $\beta$ & S24W06 \\
	2020/06/10 & 1 & 50 & $\alpha$ & S24W21 \\
	2020/06/11 & 1 & 50 & $\alpha$ & S22W35 \\
	2020/06/12 & 1 & 50 & $\alpha$ & S26W46 \\
	2020/06/13 & 1 & 60 & $\alpha$ & S25W60 \\
	2020/06/14 & 1 & 20 & $\alpha$ & S25W73 \\
	2020/06/15 & 1 & 10 & $\alpha$ & S26W86 \\
 \hline
 \end{tabular}
\end{table}

The spectrum records of solar radio emission were obtained by using the receiver DSP-Z (short abbreviation for digital spectropolarimeters of type Z) which is a standard device of the radio telescope UTR-2 (Zakharenko et al.~\cite{zakh16}). This allows performing the real-time FFT analysis in two independent channels whose operating frequency band is 0-33 MHz. Since the prototype antenna can receive radio emission up to 70 MHz, we applied a special technique for records of radio emission in the frequency band twice as wide as the receiver can perform in each channel separately. While one channel received radio signals in the frequency range 4-33 MHz, another operated within 33-66 MHz. Their combination gave the spectrum of radio emission within 4-66 MHz. Based on the technique, the frequency and time resolutions were 4 kHz and 100 ms, respectively, in each channel.

\section{Observations}

Radio emission of solar bursts is often divided into types from the analysis of their observed frequency drift rate. The most numerous among the solar bursts are bursts of type III (see, for example, Reid and Ratcliffe~\cite{rev14} with references therein). They are caused by high velocity electron beams accelerated by unstable magnetic fields of the solar atmosphere, and because of this, their frequency drift rate is extremely high. Typically, the type III bursts manifest a monotonic shape in dynamic spectra, going from high to low frequencies, showing the motion of beams to Earth. However, sometimes the closed magnetic fields on the Sun can change this exhibition in dynamic spectra. Their overall spectral signature resembles the letters J and U, and the bursts are therefore called the type J and the type U, respectively (Fokker~\cite{fokker70}, Labrum and Stewart~\cite{lab70}, Stone and Fainberg~\cite{stone71}, Caroubalos et al.~\cite{car73}, Morosan et al.~\cite{morosan17}). It is generally accepted that the origin of these bursts is explained by the fact that the beam electrons travel along closed magnetic field lines, and therefore the frequency drift velocity is inverted (Zhelezniakov~\cite{zhel69}). The motion of electrons from low to high density regions causes an increase of radio emission in frequency, whereas if the electron beam moves away from the Sun, the frequency of radio emission decreases (Leblanc et al.~\cite{leblanc83}, Aschwanden et al.~\cite{aschw92}). The exciters, going through the corona, can produce simultaneous emission at the fundamental and the second harmonic of the local coronal plasma frequency. In comparison with ordinary type III bursts both U and J bursts occur rarely. In our observations we detected a unique event in which the U-burst was surrounded not only by type III bursts, but associated with some of them being interplanetary.

Although the Sun in the summer of 2020 was at solar activity minimum, nevertheless, its activity has been still sometimes occurred. In particular, on May 29, 2020, the sunspot group caused the largest solar flare in a long period, starting from October 2017. This flare was of the M class, and it could be seen as the beginning of the awakening of the Sun after a long inactivity. Recall that the solar flare strength is denoted by letters A, B, C, M, and X with decimal sub-classes. The smallest of them are A-class, whereas X-class flares are the most intense. The M-class flare allowed us to assume continuation of solar activity in the following days and a chance for recording solar radio bursts in early June 2020. Really, a new active area, namely NOAA AR12765, appeared on the limb from the eastern side of the Sun on June 3, 2020 (see Table~\ref{tab1}). At first it consisted of a single spot (https://www.spaceweatherlive.com/en/solar-activity/region/12765), which turned into a small bipolar region on June 5, 2020, close to the north-east of the initial spot position. Generally, sunspots can have different sizes and shapes. On this day the size of sunspots was the largest for this active region. Its maximum value was 130~MH (in ``millionths of the visible solar hemisphere''). It should be mentioned that most sunspot groups cover an area, comparable to the entire surface area of the Earth (which is almost 170~MH), but  larger sunspot groups can easily reach 1000~MH or even much more (Meadows~\cite{mea02}). In the following days the region remained in the $\beta$ class, but decaying, after 10 June it moved to the $\alpha$ class until it disappeared completely, going to the far side of the Sun. Note that the B6 (faint) flare occurred in this region at $\sim$21:33 UT on June 8, 2020. 

\begin{figure}
\centerline{\includegraphics[width=13cm, angle=0]{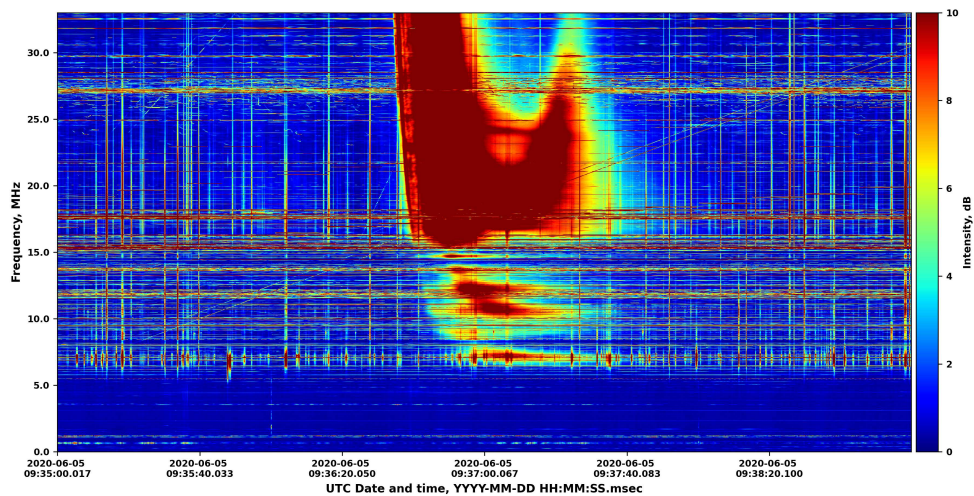}}
%\centerline{\includegraphics[width=8cm, angle=0]{moon_ant.eps}
%\includegraphics[width=5cm, angle=0]{antenna.eps}}
\caption{(Color online) Dynamic spectrum of the solar U+III bursts (observed on 5 June 2020) with help of the antenna prototype intended for ultra-long-wavelength records. The prominent narrow vertical and horizontal lines indicate numerous intensive radio disturbances generated by natural lightning discharges and broadcast stations, respectively.}
\label{fig1}
\end{figure}

\begin{figure}
\centerline{\includegraphics[width=13cm, angle=0]{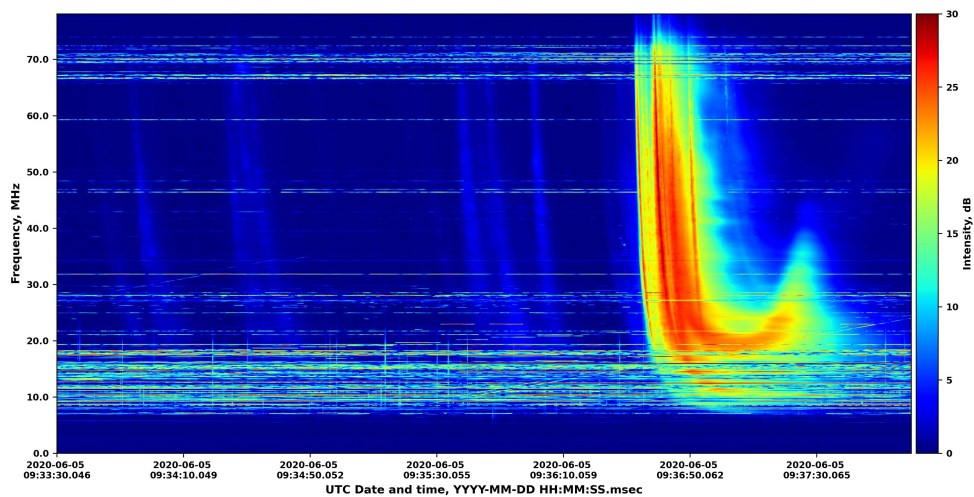}}
\caption{(Color online) Dynamic spectrum of many solar bursts at $\sim$09:37 UT on 5 June 2020 obtained with the active antenna array GURT (time resolution 100 ms, frequency resolution 38 kHz).}
\label{fig2}
\end{figure}

Bipolar magnetic fields are interesting primarily because they can be responsible for the type U solar radio bursts. It was such a burst that we managed to successfully record on June 5, 2020 (Fig.~2) with the help of the antenna prototype for future lunar radio telescopes. The change of sign in the frequency drift rate from negative to positive in this burst was clearly seen in the frequencies of $\sim$20 MHz to $\sim$15 MHz. Due to low solar activity, in general, the cutoff frequency of the Earth's ionosphere at our latitude is noticeably lower than 10 MHz (in some days even reaching frequencies of 2 MHz). This makes it possible to conduct radio observations at frequencies pretty close to those that are planned to record with lunar observatories. Therefore, the received bursts were confidently detected at extremely low frequencies close to 6 MHz, despite numerous natural and artificial radio interference. The much more detailed spectrum above 8 MHz was provided by recording this event with the GURT antenna array (Fig.~3). This active antenna consists of 25 cross-dipoles, five in each row and column. Dipole arms have the east-west and north-south orientations. The instrument serves for receiving radio emission within 8-80 MHz (Konovalenko et a.~\cite{konov16}). As the sensitivity of the GURT array is higher than one dipole antenna has, its dynamic spectrum represents many (less intense) type III solar bursts clearly visible before the solar U-burst. Moreover, the latter has a fine structure.

\begin{figure}
\centerline{\includegraphics[width=13cm, angle=0]{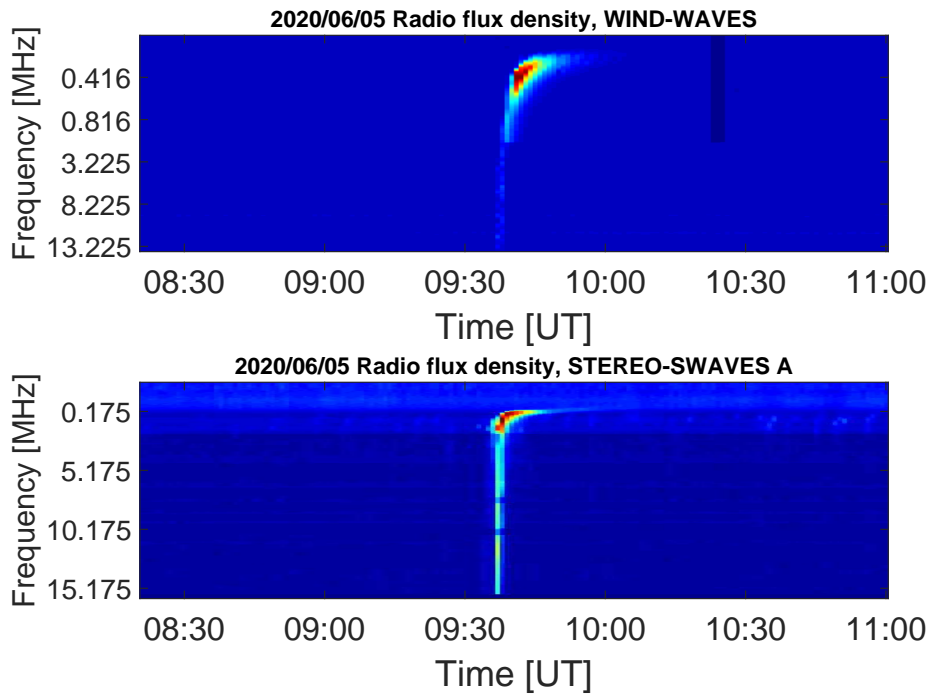}}
\caption{(Color online) Dynamic spectra of radio emission recorded with the solar space-based observatories WIND and STEREO A on 5 June of 2020, according to https://solar-radio.gsfc.nasa.gov/data/wind/rad2 and https://solar-radio.gsfc.nasa.gov/data/stereo/new\_summary.}
\label{fig3}
\end{figure}

The space-based observations with the STEREO A and the WIND spacecrafts at this time support our detection (https://solar-radio.gsfc.nasa.gov/data/WWaves-SWAVES/2020/wind\_stereo\_20200605.pdf). Unfortunately, the highest frequencies of these radio instruments ($\sim$16 MHz and $\sim$13.8 MHz, respectively) did not allow noticing all the details of this event. Nevertheless, the observations show continuation of the solar radio event up to 100 kHz which may be interpreted as an interplanetary type III burst (Fig.~4). Interestingly, this result manifests an important feature of this event. We observe the separation of electron beams into two parts. One part, due to the deflection of beams by solar magnetic fields back to the Sun, caused the emergence of the U-shaped burst, whereas another followed in the opposite direction to Earth, producing traces on the dynamic spectra, characteristic for the solar bursts of type III, merging together. The types of solar bursts, observed simultaneously, may be called the U+III bursts, since they were connected with the same place (AR12765) on the Sun and generated by the same group of associated electron beams. The U+III event is superimposed to a series of weaker type III bursts.

\begin{figure}
\centerline{\includegraphics[width=13cm, angle=0]{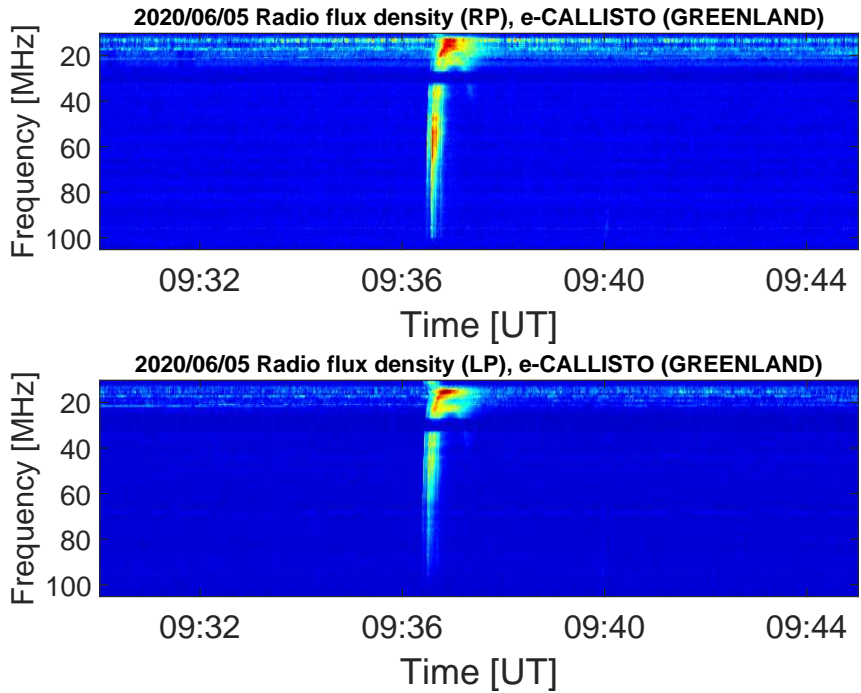}}
\caption{(Color online) Dynamic spectra of radio emission in right (RP) and left (LP) handed polarizations from the solar observations on 5 June of 2020 with the e-Callisto network station in Greenland. In the upper part of the spectra, the track similar to a type III burst can be seen, separating from the type U burst branch with the negative frequency drift rate.}
\label{fig4}
\end{figure}

To confirm this concept, we have found the U-shaped burst in the observations made on other suitable radio telescopes, which had a similar low-frequency range for recording solar radio emission. One of e-Callisto network stations (Benz et al.~\cite{benz09}), located in Greenland (ISR Kellyville observatory near Kangerlussuaq, 66$^\circ$ 59$^\prime$ 8.88$^{\prime\prime}$ N, 50$^\circ$ 56$^\prime$ 44.26$^{\prime\prime}$ W), recorded the solar U-shaped burst at the same time that we did (see Table~\ref{tab2}). Let us mention a little about hardware features of this station for solar observations: the frequency domain of 10 to 95 MHz is divided into 200 frequency channels each of which has the band of 475 kHz, and the time resolution is 250 ms. The antenna consists of one broadband low-frequency antenna intended for the Long Wavelength Array (LWA, see http://lwa.phys.unm.edu). The spectra obtained by the e-Callisto station (see Fig.~5) are morphologically identical to the spectrum obtained by our instruments, which had noticeably both better sensitivity and higher frequency-time resolution. There is no doubt that the records show the same event, manifesting the solar U-shaped burst. Moreover, fragments of this event was also detected with other e-Callisto stations. It is also worth mentioning that this group of solar bursts was recorded by the Nan\c{c}ay Decametric Array (NDA) in France. The dynamic spectrum is shown in Fig.~6. The radio telescope provides daily observations of solar radio emission at 10-80 MHz. Its 2 $\times$ 72 helical spiral antennas are for polarization measurements (Boischot et al.~\cite{nda}), and the solar NDA data are freely available for viewing (via the link https://realtime.obs-nancay.fr). Using the variety of radio astronomy tools in our comparison, we conclude that the observed U+III bursts have the solar origin and cannot be associated with ionospheric disturbances and/or equipment malfunctions.

\begin{table}
 \centering
 \caption{High frequency cutoff for the U+III bursts, observed on June 5, 2020 at $\sim$09:37 UT, from the e-Callisto network data  (see http://soleil.i4ds.ch/solarradio/callistoQuicklooks/ ?date=20200605).}\label{tab2}
 \vspace*{1ex}
 \begin{tabular}{|c|c|c|c|}
  \hline
   Country & Location & Frequency & Cutoff, \\
	         &          & range, MHz &   MHz  \\
  \hline
  Denmark & Copenhagen, DTU & 45-100 & 85 \\  
  United Kingdom & Glasgow, UOG & 45-81 & $>$ 81 \\
	Switzerland  & Landschlacht, HB9SCT	 & 15-85 & 82 \\
	Switzerland  & Landschlacht, HB9SCT & 45-155 & 84 \\
	Switzerland  & Heiterswil, Kant & 45-81 & $>$ 81  \\
	Austria			 & Michelbach, Lensch & 20-90 & 75  \\
	Belgium      & Humain, ROB & 45-435 & 80-90  \\
	India 		   & Bangalore, GAURI & 45-410 & 85  \\
	Denmark 		 & Kangarlussuaq, Greenland & 10-95 & 85-90  \\
 \hline
 \end{tabular}
\end{table}

\section{Results}

According to Reid and Kontar~(\cite{kontar17}), the generation of radio emission in the form of the solar U bursts simultaneously with type III bursts requires strictly defined conditions. Initially, having a negative velocity gradient, an injected electron beam cannot generate radio emission because of its stable distribution function to Langmuir wave production. Nevertheless, when it propagates through solar corona, a positive velocity gradient becomes its feature, making the beam unstable and producing a solar burst. The propagation effects determine a starting height with which the burst is nascent. The value is 0.6$R_\odot$ (from the solar photosphere). Although it is dependent on parameters of beams, its value may be taken as an estimate. If one assumes that this conjecture is true, with its help we can calibrate the electron density model of solar corona (see more details below). Before 0.6$R_\odot$ an electron beam does not produce radio emission, but after that three regimes are possible. The electron beam with too low density still does not emit. In the second regime, for moderate densities, a beam travels along open magnetic field lines, producing only type III bursts. The third regime, the most interesting for us, is characterized by high initial electron beam densities, that is good for uprising both type III and type U bursts in radio emission. In this case the magnetic loop should be large enough for propagation effects, making the electron beam unstable, until it reaches the loop top and turns back to the Sun. This does not exclude the generation of type III bursts, due to the movement of beams along open magnetic field lines. As a rule, radio emission of type U bursts has a weaker and more diffuse shape for the positive frequency drift rate branch (Aurass and Klein~\cite{ak97}). The fine structure of the U+III bursts manifests many radio sources tracing a similar path through the corona so that each of the radio patterns was produced by its electron beam.  

\begin{figure}
\centerline{\includegraphics[width=13cm, angle=0]{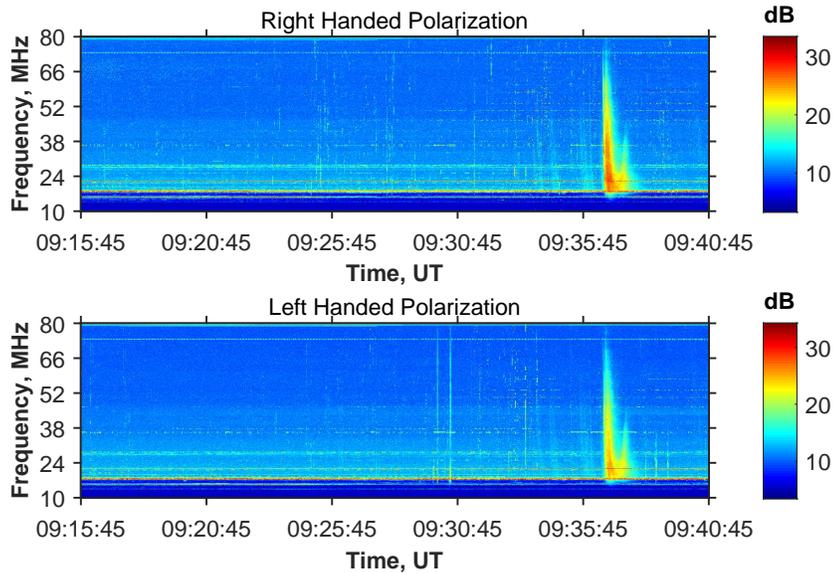}}
\caption{(Color online) Dynamic spectra of solar radio emission on 5 June of 2020 observed with NDA in Nan\c{c}ay.}
\label{fig5}
\end{figure}

Using the Newkirk (one-fold) model for the solar corona (Newkirk~\cite{newkirk}), the radial change of the electron density is written as
\begin{equation}
n_e(r)=\alpha\cdot 4.2\cdot 10^4\cdot\,10^{(4.32/r)}\,,
\end{equation}
where $n_e(r)$ is the electron density of coronal plasma in cm$^{-3}$, depending on the heliocentric height $r$ in solar radii. This $\alpha$-fold model well describes the electron density profile above quiet equatorial regions ($\alpha$ = 1), in dense loops ($\alpha$ = 4), and in extremely dense loops for $\alpha$ = 10 (Koutchmy~\cite{koutchmy}, Mann et al.~\cite{mann18}), but the value of the parameter $\alpha$ is indicative. As the local plasma frequency $f_{pe}$ (in MHz) equals
\begin{equation}
f_{pe}=8.9\cdot 10^{-3}\sqrt{n_e}\,,
\end{equation}
then the source height and the radiation frequency are linked by a simple expression
\begin{equation}
r=\frac{2.16}{\lg f - 0.26 - 0.5\lg\alpha}\,,
\end{equation}
where $f$ is the radiation frequency in MHz. As applied to our event and following Table~\ref{tab2}, the U+III bursts started with $\sim$85 MHz that corresponds to the heliocentric height $\sim$1.6$R_\odot$ (or $\sim$0.6$R_\odot$ from photoshere) under $\alpha\approx 4$. This is in good agreement with conjectures of Reid and Kontar~(\cite{kontar17}). Note that the high-frequency cutoff finding allows calibrating the value of $\alpha$ for the coronal density model. The top of the U-burst reached $\sim$15 MHz, i.\,e. the height was about 3.5$R_\odot$, taking $\alpha\approx 4$. In the pivot point the instantaneous bandwidth was about 5 MHz, permitting us to estimate the loop width at the top. It was about 0.6$R_\odot$. Recall that the activity of solar regions depends of their size (Solanki~\cite{solanki03}). The larger it is, the higher the probability of occurrence of flares, coronal mass ejections, diversity of solar bursts and their intensity. The active region 12765 was developed from the unipolar form to bipolar, and we observed the only U-burst, when the size of this region was maximum. Thus, one can assume that each bipolar active region with the size more 110~MH will be able to generate electron beams with high initial densities, resulting in radio emission of the U-bursts simultaneously and together with type III bursts. Finally, it should be noticed that according to Fig.~3 and Fig.~6, the U+III bursts are surrounded by weaker type III bursts which were caused by electron beams with moderate densities. This case can correspond to the second regime mentioned above.

\section{Conclusions}

Having such an extensive set of radio and optic observations on 5 June 2020, we can confidently assert that the event at $\sim$09:37 UT included the solar U-burst combined with type III bursts. Evidently, they were generated by many electron beams accelerated in the active region 12765 and traveling in the solar corona along both close and open magnetic field lines. The uniqueness of this phenomenon lies in the fact that magnetic fields in AR12765 were developed only between a unipolar ($\alpha$ class) and a bipolar configuration ($\beta$ class). This was due to low solar activity. The configuration of magnetic fields and the size of the active region were favorable to the emergence of solar bursts of such types. Observations with help of various radio instruments allowed us to explore this event in more detail and to get reliable results, confirming the model of this phenomenon. Moreover, this research also gave a number of new interesting footings related to the development of ultra-long-wavelength antennas for future lunar radio telescopes.\\

\begin{acknowledgements}
The authors thank the WIND, STEREO, NDA and e-Callisto network teams for their instrument maintenance and open data access. This research was partially supported by Research Grant  0120U101334, and authors acknowledge the National Academy of Sciences of Ukraine for this support. The authors are also thankful to Dorovskyy V.V. for helpful discussions.
\end{acknowledgements}

\label{lastpage}

\end{document}